\documentclass[aps,prl,final,showpacs,nobibnotes,floatfix]{revtex4}

\usepackage[latin1]{inputenc}                    
\usepackage{graphicx}                            
\usepackage{latexsym}                            
\usepackage{amsfonts}                            
\usepackage{amssymb}                             
\usepackage[mathscr]{eucal}                      
\usepackage{dcolumn}                             
\usepackage{theorem}                             
\usepackage{epsfig}                              
\usepackage{amsmath}                              
\usepackage{longtable}
\usepackage{wasysym}

\begin{document}

\begin{flushleft}
DESY 04-169\\
Edinburgh 2004/16\\
LU-ITP 2004/027\\
LTH 633
\end{flushleft}

\title{Determination of Light and Strange Quark Masses from Full
  Lattice QCD} 

\author{M. G\"ockeler$^{1,2}$, R. Horsley$^{3}$, A.C. Irving$^{4}$,
        D. Pleiter$^{5}$, 
        P.E.L. Rakow$^{4}$, G. Schierholz$^{5,6}$ and H. St\"uben$^{7}$
\vspace {0.15cm}}

\affiliation{$^1$ Institut f\"ur Theoretische Physik, Universit\"at Leipzig, 
D-04109 Leipzig, Germany\\
$^2$ Institut f\"ur Theoretische Physik, Universit\"at Regensburg, 
D-93040 Regensburg, Germany\\
$^3$ School of Physics, University of Edinburgh, Edinburgh EH9 3JZ, UK\\
$^4$ Theoretical Physics Division, Department of Mathematical 
Sciences, University of Liverpool, Liverpool L69 3BX, UK\\
$^5$ John von Neumann-Institut f\"ur Computing NIC, Deutsches 
Elektronen-Synchrotron DESY, D-15738 Zeuthen, Germany\\
$^6$ Deutsches Elektronen-Synchrotron DESY, D-22603 Hamburg, Germany\\
$^7$ Konrad-Zuse-Zentrum f\"ur Informationstechnik Berlin ZIB, D-14195 Berlin,
Germany}
\author{\vspace*{-0.3cm} - QCDSF--UKQCD Collaboration -}
\noaffiliation

\begin{abstract}
We compute the light and strange quark masses $m_\ell = (m_u+m_d)/2$ and 
$m_s$, respectively, in full lattice QCD with $N_f=2$ flavors of light
dynamical quarks. The renormalization constants, which convert bare quark 
masses into renormalized quark masses, are computed nonperturbatively,
including the effect of quark-line disconnected diagrams. 
We obtain $m_\ell^{\overline{MS}}(2\,\mbox{GeV})=4.7(2)(3)
\,\mbox{MeV}$ and $m_s^{\overline{MS}}(2\,\mbox{GeV})=119(5)(8) \,\mbox{MeV}$.

\end{abstract}


\maketitle

The light and strange quark masses are among the least well known parameters
of the Standard Model. The reason is that quarks are confined, so that the
masses must be determined indirectly through their influence on hadronic
observables. This requires nonperturbative techniques. One such technique
is lattice QCD.  

The quark masses obtained directly in lattice calculations are bare quark
masses at the cut-off scale $a^{-1}$, where $a$ denotes the lattice spacing.
For the lattice numbers to be useful for phenomenology, it is necessary to
convert the bare quark masses to renormalized masses in some standard
renormalization scheme. Because lattice perturbation theory converges badly,
and the expansion coefficients are generally known to one loop order only,
this ought to be done nonperturbatively. In full QCD a one-loop perturbative 
renormalization of the mass operator is totally inadequate even, as it does
not account for the disconnected (flavor singlet) contribution shown in Fig.~1, which turns
out to be comparable with the connected contribution at present lattice
spacings. 

\begin{figure}[b]
\begin{center}
\epsfig{file=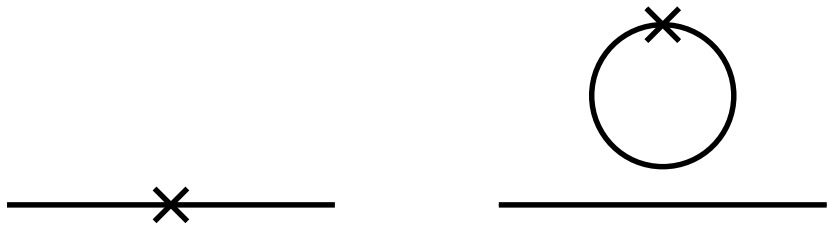,width=3cm,angle=270,clip=}
\end{center}
\caption{Quark diagrams contributing to the renormalization of the mass
operator ($\mathbf{\times}$). The left figure shows the connected (nonsinglet)
contribution, the right figure the disconnected (singlet minus nonsinglet) 
contribution. Gluon lines have been omitted.} 
\end{figure}

In quenched QCD, in which the effect of sea quarks is neglected
(and hence quark-line disconnected fermion loops are absent), several
groups~\cite{quenchednpt} 
have carried out an entirely nonperturbative calculation of the light and
strange quark masses. Remarkably consistent results have been found. 
Previous calculations in full QCD, both with $N_f=2$~\cite{dyn2} and
$N_f=3$~\cite{dyn3} flavors of sea quarks, employ perturbative renormalization 
to compute the relation between the bare and renormalized quark
masses. These authors, except perhaps Eicker {\it et al.}, find rather small
values for the strange quark mass, which lie substantially below the central
value quoted by the Particle Data Group~\cite{pdg}. 

\begin{figure}[b]
\vspace*{-2.25cm}
\begin{center}
\epsfig{file=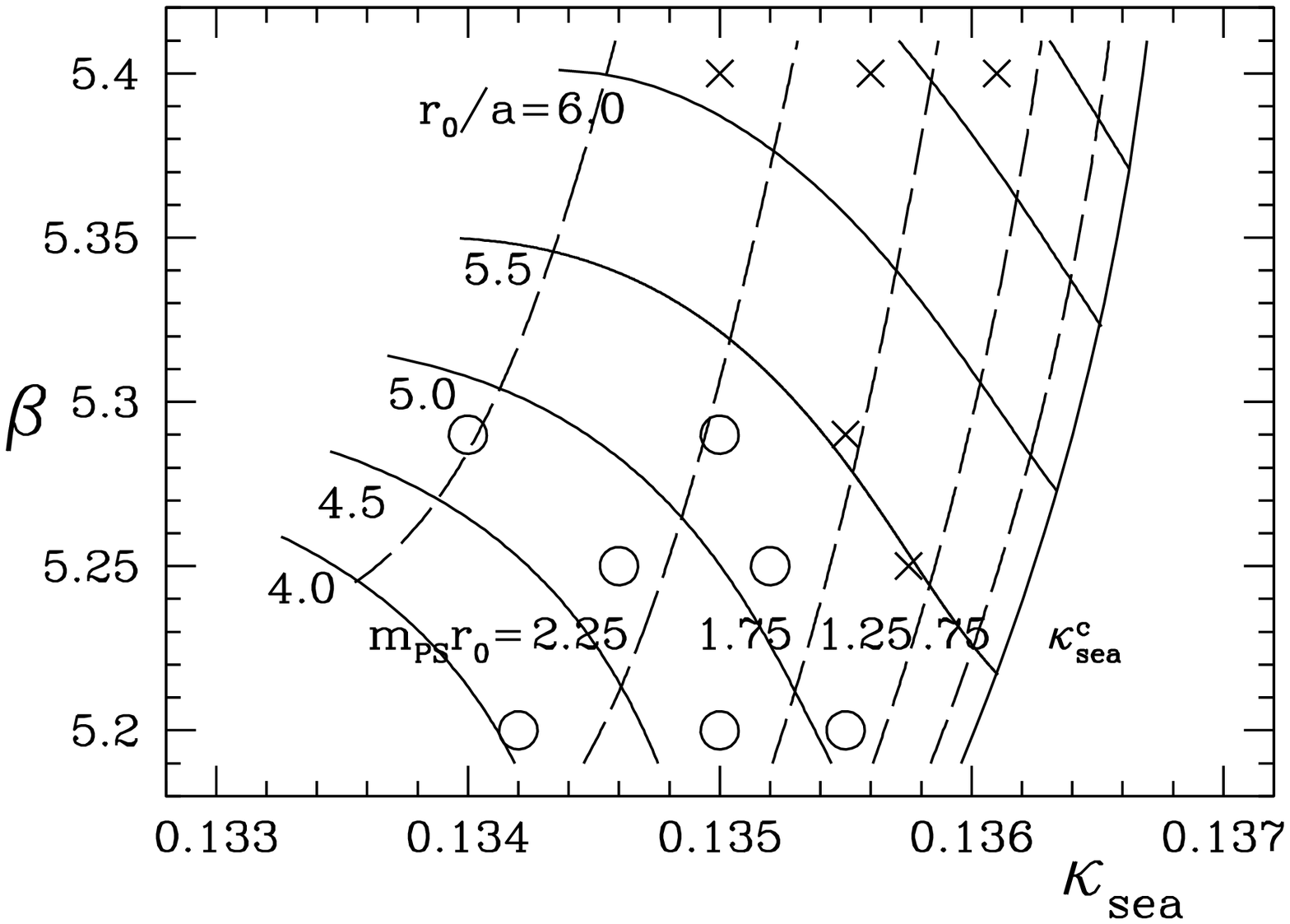,width=12.2cm,clip=}
\end{center}
\vspace*{-0.15cm}
\caption{Parameters of our dynamical gauge field configurations, together with
  lines of constant $r_0/a$ (solid lines) and lines of constant $m_{PS} r_0$
  (dashed lines). The simulations are done on $24^3\, 48$ ($\times$) and
  $16^3\, 32$ ($\Circle$) lattices, respectively.} 
\vspace*{-1.75cm}
\begin{center}\hspace*{0.35cm}
\epsfig{file=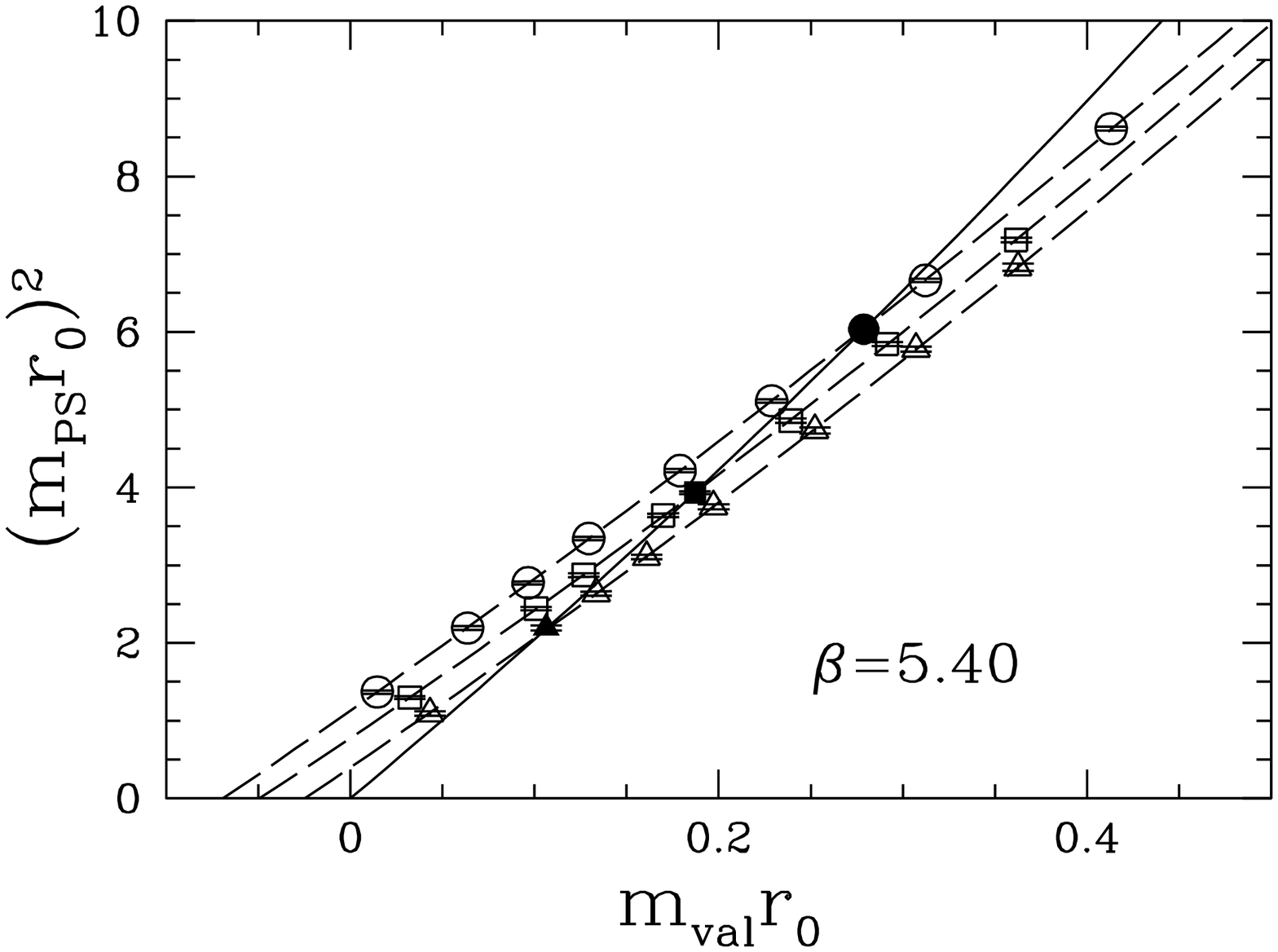,width=11.8cm,clip=}
\end{center} 
\vspace*{-0.15cm}
\caption{The partially quenched pseudoscalar mass as a function of $m_{\rm
    val}$ at $\beta=5.40$ for $\kappa_{\rm sea}=0.1350$ ($\Circle$), 
    0.1356 ($\Box$) and 0.1361 ($\triangle$), together with the fit
    (\ref{fit}). The     solid line and symbols refer to the case $\kappa_{\rm
    val} = \kappa_{\rm sea}$.}   
\end{figure}

\begin{table}[t]
\caption{Hopping parameters of sea and valence quarks used in this
  calculation, together with their critical values.} 
\begin{ruledtabular}
\begin{tabular}{c|c|ccccccccc|c|c}
$\beta$ & $\kappa_{\rm sea}$ & \multicolumn{9}{c|}{$\kappa_{\rm val}$} &
$\kappa_{\rm val}^c$ & $\kappa_{\rm sea}^c$ \\ \hline
      & 0.13420 & 0.13340 & 0.13380 & 0.13420 & 0.13470 & 0.13530 & 0.13560 &
        0.1360 & 0.13620 & & 0.137550(49) &  \\
5.20  & 0.13500 &  0.13320 & 0.13370 & 0.13420 & 0.13450 & 0.13500 & 0.13530 &
        0.1355 & 0.13570 & & 0.136889(32) & 0.136008(15) \\ 
        & 0.13550 & 0.13320 & 0.13360 & 0.13400 & 0.13430 & 0.13480 & 0.13500 &
        0.13530 & 0.13550 & 0.13570 & 0.136457(23) &  \\ \hline
      & 0.13460 &  0.13370 & 0.13400 & 0.13460 & 0.13490 & 0.13530 & 0.13550 &
        0.13590 & 0.13610 & & 0.137237(19) &  \\
5.25  & 0.13520 &  0.13370 & 0.13410 & 0.13450 & 0.13480 & 0.13520 & 0.13550 &
        0.13580 & 0.13590 & & 0.136883(13) & 0.136250(7)\phantom{9} \\ 
        & 0.13575 & 0.13360 & 0.13390 & 0.13430 & 0.13460 & 0.13500 & 0.13520
        & 0.13550 & 0.13575 & 0.13600  & 0.136553(9)\phantom{9} &  \\ \hline
      & 0.13400 &  0.13400 & 0.13440 & 0.13490 & 0.13520 & 0.13550 & 0.13570 &
        0.13590 & 0.13620 & & 0.137516(33) &  \\
5.29  & 0.13500 &  0.13400 & 0.13430 & 0.13470 & 0.13500 & 0.13550 & 0.13570 &
        0.13600 & 0.13610 & & 0.137045(16) & 0.136410(9)\phantom{9}\\ 
        & 0.13550 & 0.13390 & 0.13430 & 0.13460 & 0.13490 & 0.13530 & 0.13550 &
        0.13580 & 0.13600 & 0.13630 & 0.136816(11) &  \\ \hline
      & 0.13500 &  0.13420 & 0.13480 & 0.13500 & 0.13530 & 0.13560 & 0.13590 &
        0.13610 & 0.13630 & 0.13660 & 0.137131(14) &  \\
5.40  & 0.13560 &  0.13460 & 0.13500 & 0.13530 & 0.13560 & 0.13570 & 0.13595 &
        0.13610 & 0.13650 & & 0.136966(12) & 0.136690(22) \\ 
        & 0.13610 & 0.13470 & 0.13500 & 0.13530 & 0.13560 & 0.13580 & 0.13595 &
        0.13610 & 0.13645 & & 0.136836(14) &  \\ 

\end{tabular}
\end{ruledtabular}
\end{table}

In this Letter we shall present a first fully nonperturbative
calculation of the light and strange quark masses in full QCD, including the
effect of flavor singlet renormalization factors. We consider nonperturbatively
$O(a)$ improved Wilson fermions with $N_f=2$ flavors of degenerate dynamical
quarks and 
the Wilson gauge field action. The calculation is done in two steps. We
simulate dynamical gauge field configurations at four different values of the
coupling, $\beta$, and at three different sea quark masses each. The latter
are specified by the hopping parameter $\kappa_{\rm  sea}$. The actual
parameters, as well as the corresponding lattice spacings and pseudoscalar
mass values, are shown in Fig.~2. We use the force parameter $r_0$ to set the
scale.  
On these configurations we then perform a partially quenched
calculation of the pseudoscalar  mass, allowing for different sea and valence
quark masses, from which we derive the physical quark masses. In Table~1 we
list the hopping parameters of valence ($\kappa_{\rm  val}$) and sea quarks
considered in this calculation.  

\begin{figure}[t]
\vspace*{-2.25cm}
\begin{center}
\epsfig{file=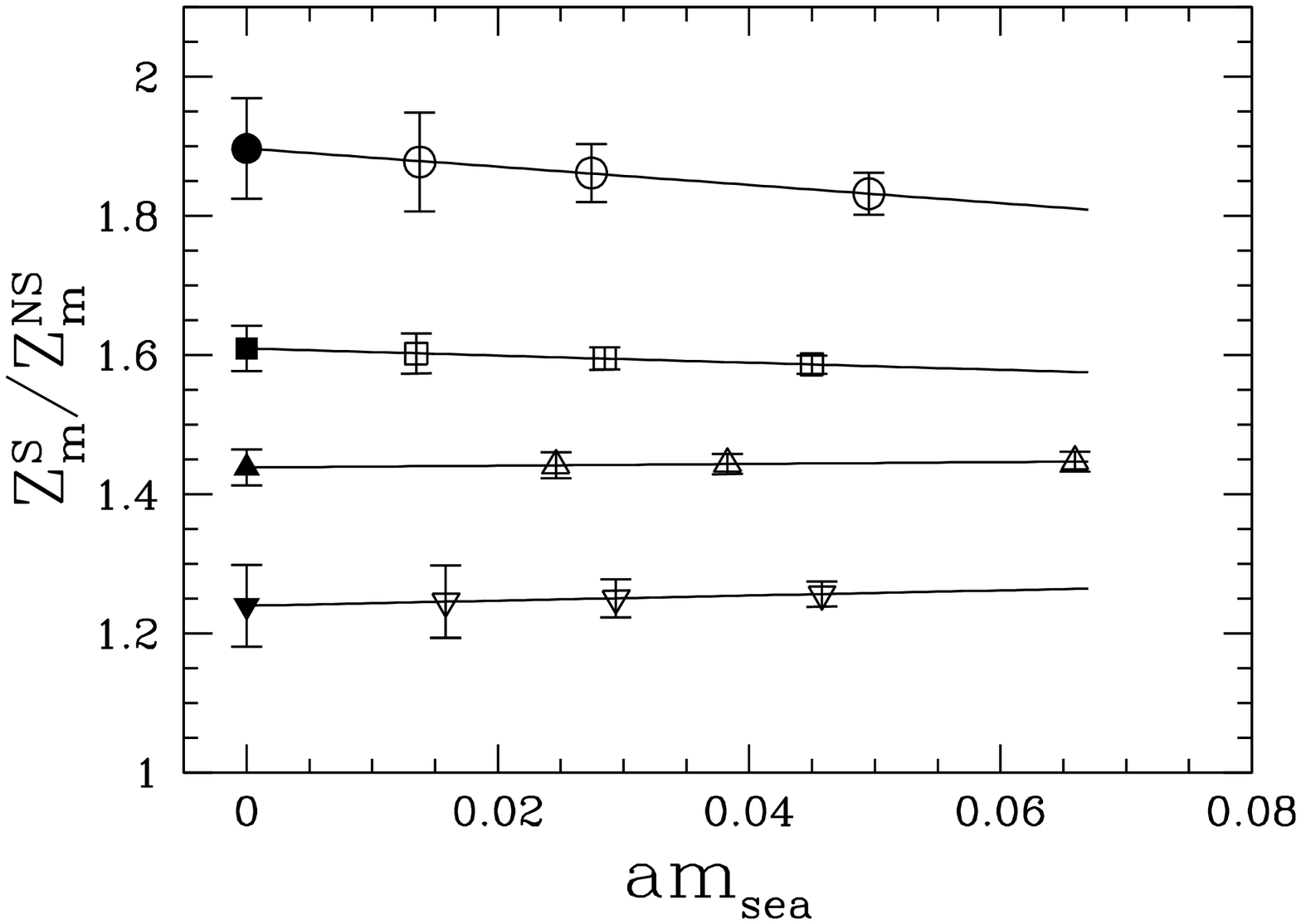,width=12cm,clip=}
\end{center}
\vspace*{-0.15cm}
\caption{The ratio $Z_m^S/Z_m^{NS}$ at $\beta=5.20$, 5.25, 5.29 and 5.40 (from
  top to bottom), together with a linear extrapolation to the chiral limit.} 
\vspace*{-1.75cm}
\begin{center}
\epsfig{file=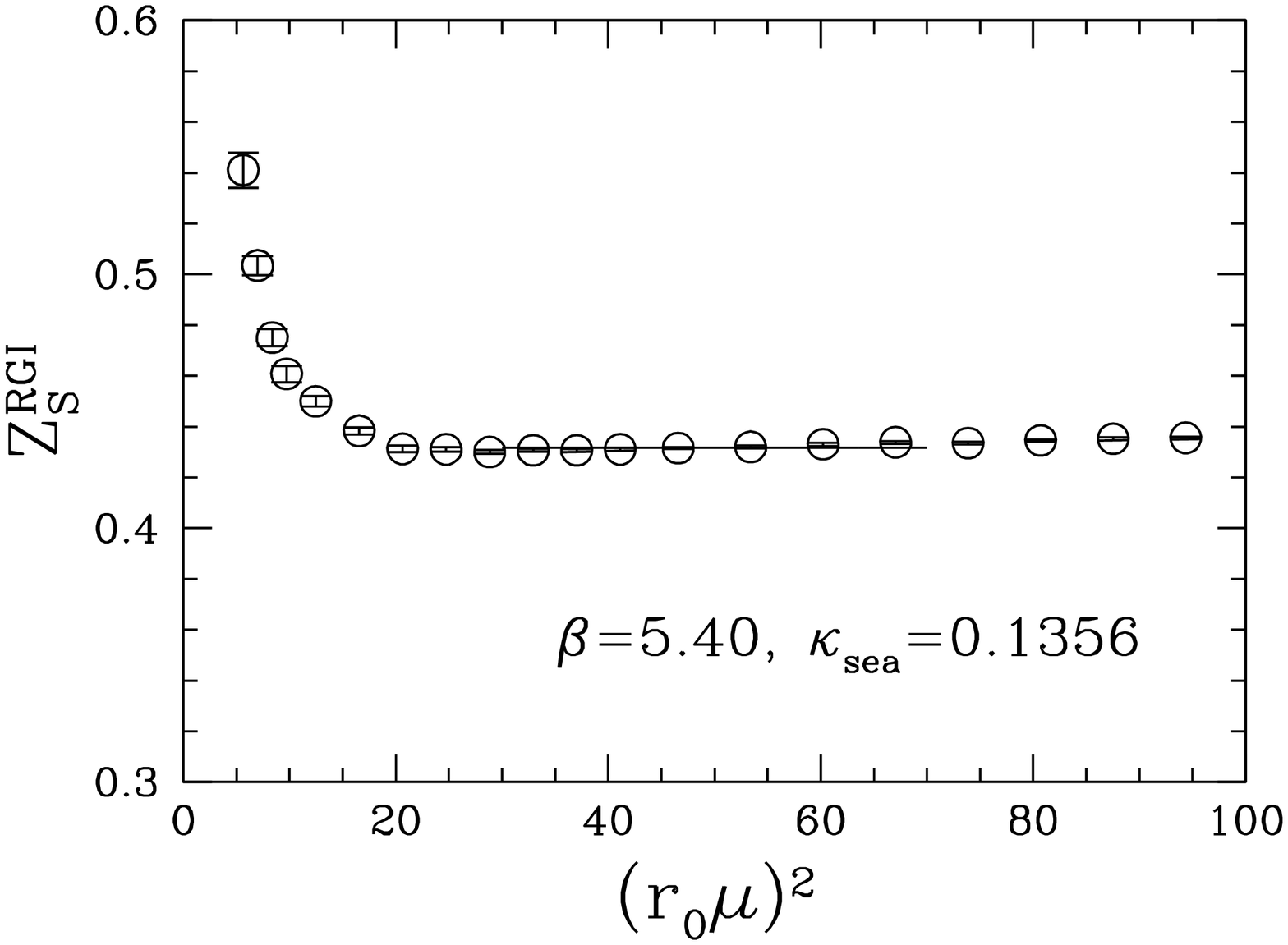,width=12cm,clip=}
\end{center}
\vspace*{-0.15cm}
\caption{The nonsinglet renormalization group invariant $Z_S^{RGI}$ at
  $\beta=5.40$, 
  $\kappa_{\rm sea} = 0.1356$ as a function of the renormalization scale
  $\mu$, together with the fit to the plateau.}  
\end{figure}

The bare sea and valence quark masses are given by $a m_{\rm sea} =
1/(2\kappa_{\rm sea})-1/(2\kappa_{\rm sea}^c)$ and $a m_{\rm val} =
1/(2\kappa_{\rm val})-1/(2\kappa_{\rm sea}^c)$, respectively. We consider the
case of degenerate valence quarks only. In Fig.~3 we show the partially
quenched pseudoscalar mass 
$m_{PS}(\kappa_{\rm sea},\kappa_{\rm val})$. The critical hopping parameter 
$\kappa_{\rm sea}^c$ is found by keeping $\beta$ fixed and varying
$\kappa_{\rm sea}$ until $m_{PS}(\kappa_{\rm sea},\kappa_{\rm sea})=0$. 
Similarly, we introduce a critical hopping parameter of the valence quarks,
$\kappa_{\rm val}^c$, which is found by varying $\kappa_{\rm val}$ until
$m_{PS}(\kappa_{\rm sea},\kappa_{\rm val})=0$, while keeping $\beta$,
$\kappa_{\rm sea}$ fixed. Our calculation requires a precise determination of
$\kappa_{\rm sea}^c$ and $\kappa_{\rm val}^c$. We perform a global fit of the
form 
\begin{equation}
\begin{split}
a r_0 m_{PS}^2 &=  
u\left(\frac{1}{\kappa_{\rm sea}} - \frac{1}{\kappa_{\rm sea}^c}\right) 
+ v\left(\frac{1}{\kappa_{\rm val}} - \frac{1}{\kappa_{\rm sea}}\right) 
+\! w\left(\frac{1}{\kappa_{\rm sea}} - \frac{1}{\kappa_{\rm
sea}^c}\right)^2\\  
&\,+ x\left(\frac{1}{\kappa_{\rm val}} - \frac{1}{\kappa_{\rm sea}}\right)^2 
+ y\left(\frac{1}{\kappa_{\rm sea}} - \frac{1}{\kappa_{\rm sea}^c}\right)
\left(\frac{1}{\kappa_{\rm val}} - \frac{1}{\kappa_{\rm sea}}\right)
\end{split}
\label{fit}
\end{equation}
to all 100 data points,
where the parameters are taken to be second order polynomials in $\beta$.  
The fit gave $\chi^2/{\rm dof} = 0.67$. The resulting values of $\kappa_{\rm
sea}^c$ and  $\kappa_{\rm val}^c$ are given in Table~1. The force parameter
$r_0/a$ was computed from the static potential.  

Taking the derivatives of the quark propagator with respect to $am_{\rm sea}$
and $am_{\rm val}$, we obtain for the renormalized sea and valence quark 
masses~\cite{tbp}
\begin{eqnarray}
m_{\rm sea}^R &=& Z_m^S m_{\rm sea}\,, \\
m_{\rm val}^R &=& Z_m^{NS} (m_{\rm val}-m_{\rm sea}) + Z_m^S m_{\rm sea}\,,
\label{z}
\end{eqnarray}
where $Z_m^S$ and $Z_m^{NS}$ are singlet and nonsinglet renormalization
constants of the mass operator. 
Partially
quenched chiral perturbation theory to NLO predicts~\cite{pqcpt}
\begin{equation}
m_{PS}^2 = [A + (B + C\ln m_{\rm val}^R)\, m_{\rm sea}^R]\, m_{\rm val}^R
          + (D + E\ln m_{\rm val}^R)\, (m_{\rm val}^R)^2 \,.
\label{chiral}
\end{equation} 
From (\ref{chiral}) follows that $m_{\rm val}^R$ vanishes where the partially
quenched 
pseudoscalar mass vanishes, which happens at the value $\kappa_{\rm 
val} = \kappa_{\rm val}^c$. If we insert this value into (\ref{z}) we obtain
the ratio
\begin{equation}
\frac{Z_m^S}{Z_m^{NS}} = \left.\frac{m_{\rm sea} - m_{\rm val}}{m_{\rm
    sea}}\right|_{\kappa_{\rm val}=\kappa_{\rm val}^c} 
    =
    \left(\frac{1}{2\kappa_{\rm sea}}-\frac{1}{2\kappa_{\rm val}^c}\right) 
    \left(\frac{1}{2\kappa_{\rm sea}}-\frac{1}{2\kappa_{\rm
    sea}^c}\right)^{-1} \,.
\end{equation}
In Fig.~4 we plot $Z_m^S/Z_m^{NS}$ for all data sets. The effect of the
quark-line disconnected diagram is found to be significant. The numbers depend
only mildly on the sea quark mass.

\begin{figure}[t]
\vspace*{-2.25cm}
\begin{center}
\epsfig{file=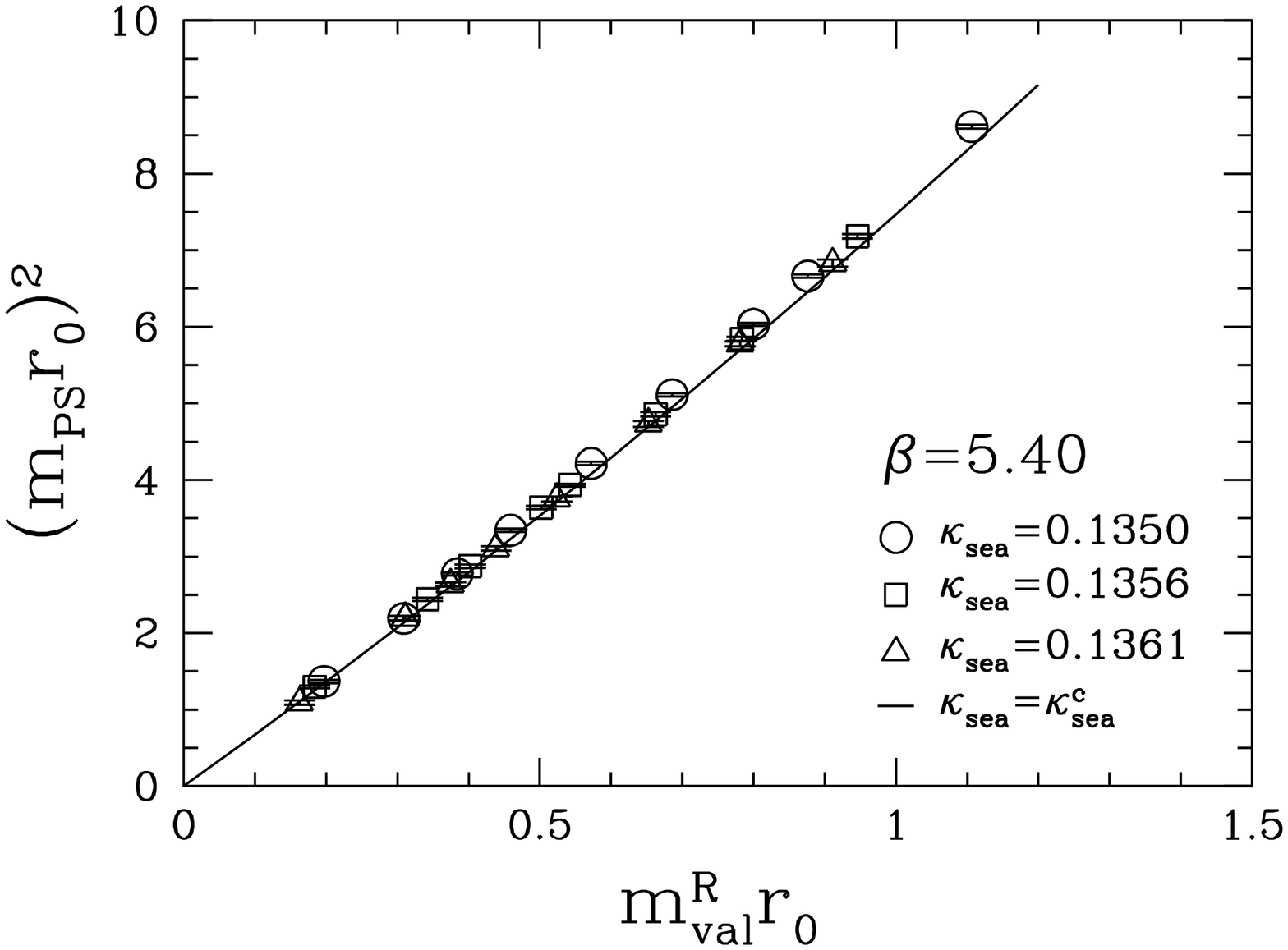,width=12cm,clip=}
\end{center}
\vspace*{-0.15cm}
\caption{The partially quenched pseudoscalar mass $m_{PS}$ as a function of
  $m_{\rm val}^R$ at $\beta=5.40$ for our three sea quark masses. The solid
  line shows the result of the fit for $m_{\rm  sea}^R=0$.}  
\vspace*{-1.75cm}
\begin{center}
\epsfig{file=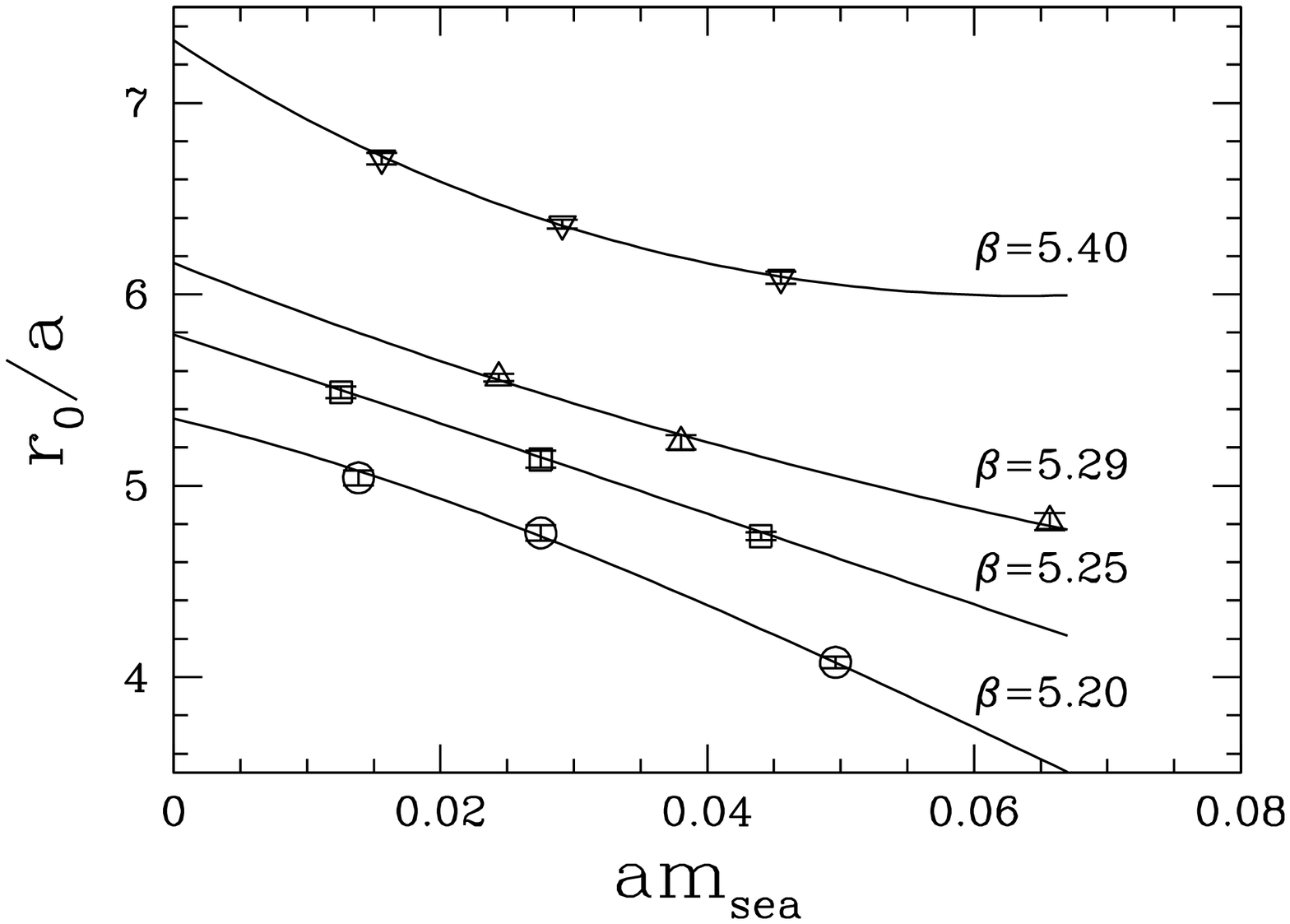,width=12cm,clip=}
\end{center}
\vspace*{-0.15cm}
\caption{The lattice spacing $a$ extrapolated to the chiral limit.}  
\end{figure}

\begin{figure}[b]
\vspace*{-5cm}
\begin{center}
\epsfig{file=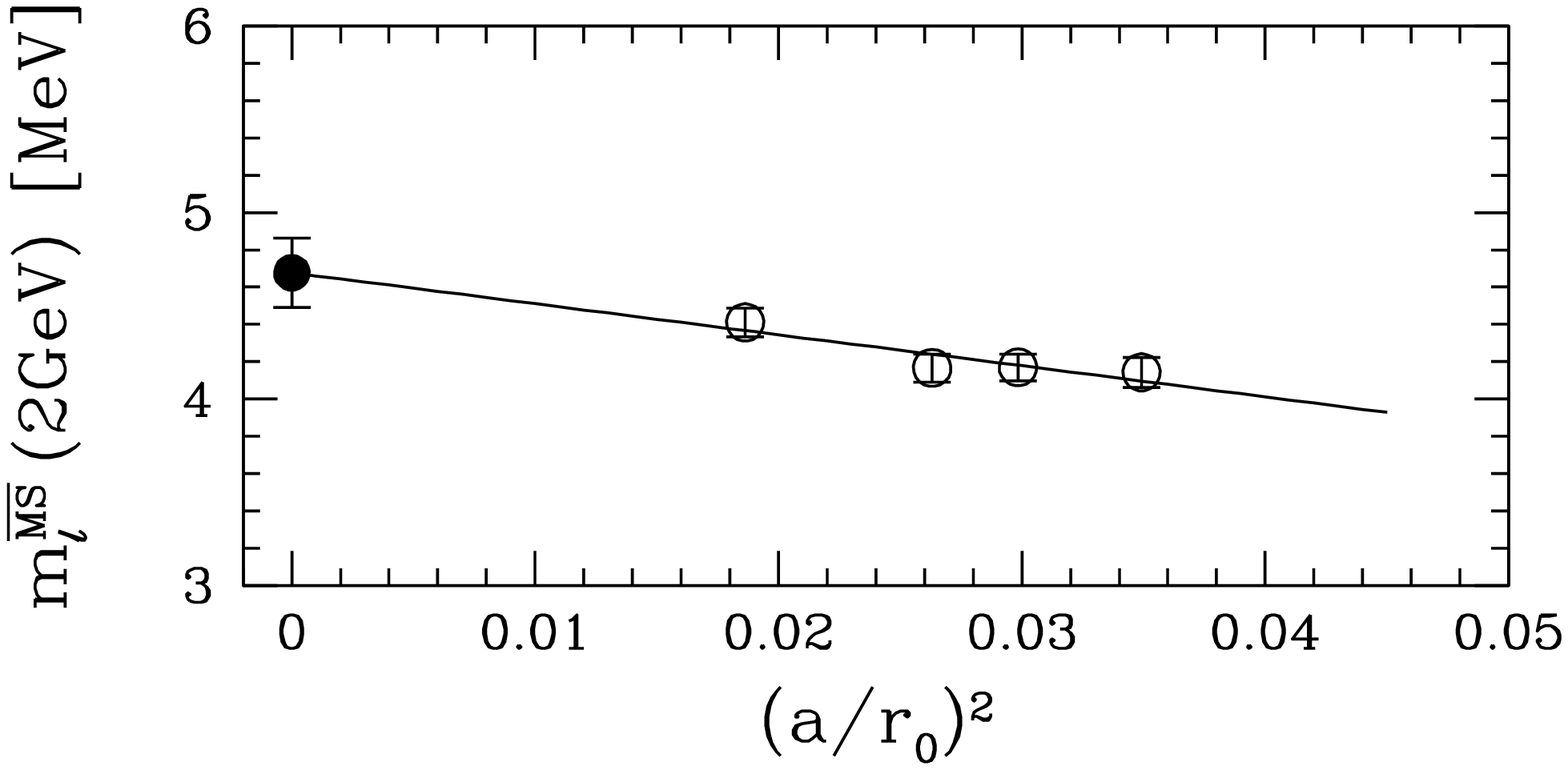,width=12cm,clip=}\\[-6cm]
\epsfig{file=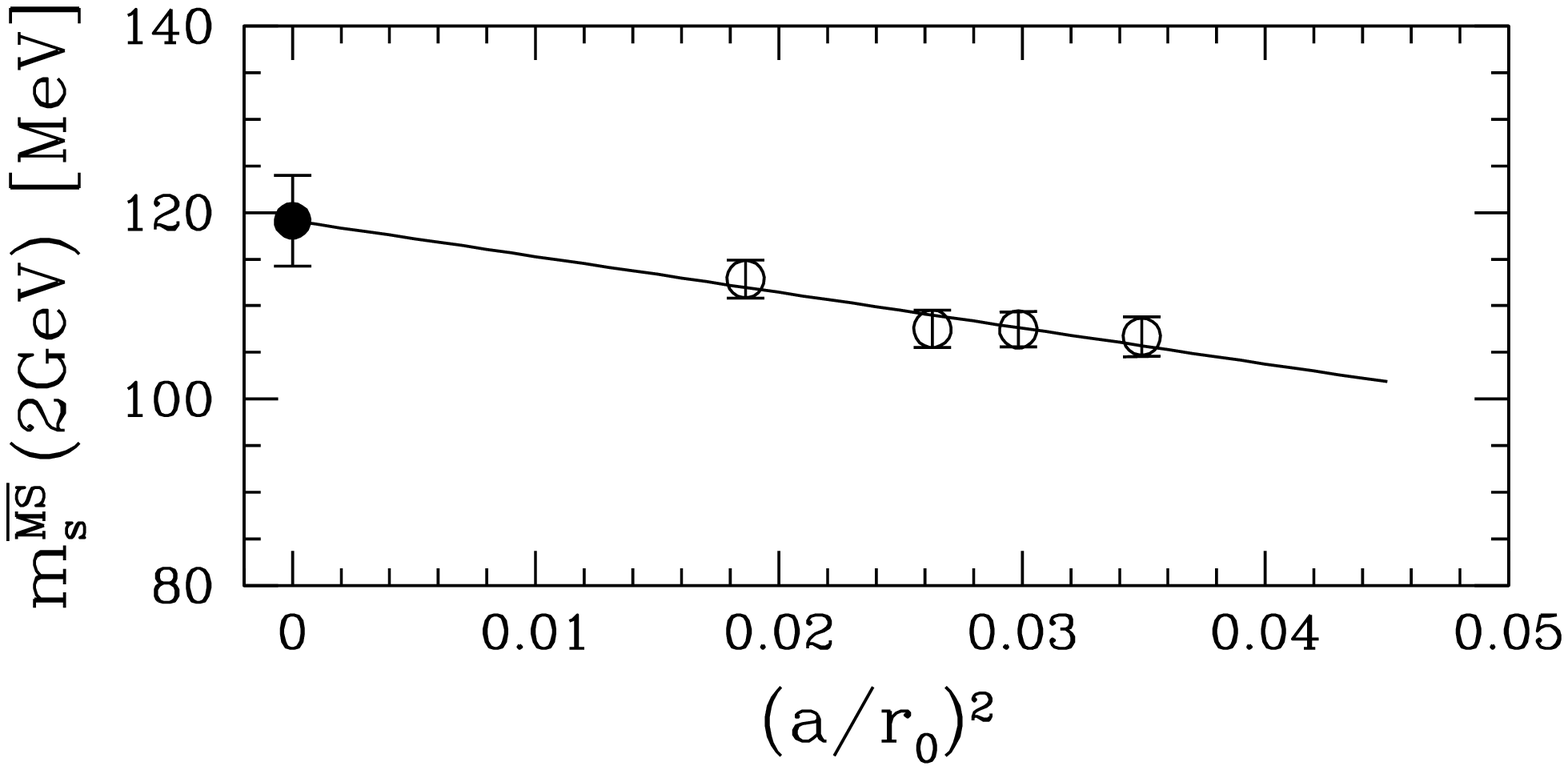,width=12cm,clip=}
\end{center}
\vspace*{-0.15cm}
\caption{The light and strange quark masses, together with the extrapolation
  to the continuum limit. The errors shown are statistical only.}   
\end{figure}

It remains to determine $Z_m^{NS}$. We compute $Z_S^{NS} = (Z_m^{NS})^{-1}$
nonperturbatively~\cite{m&us} in the {\em RI-MOM} scheme. The result is
converted to the more popular $\overline{MS}$ and {\em RGI} schemes by a
three-loop perturbative calculation~\cite{Che}. In Fig.~5 we show the
nonsinglet $Z_S^{RGI}$ as a function of the renormalization scale $\mu$. We
find that the 
nonperturbative scale dependence of $Z_S^{RI\!-\!MOM}$ is matched by the
three-loop conversion 
factor for $(r_0 \mu)^2 \gtrsim 20$. We obtain $Z_S^{RGI}$ from a fit to the
plateau as indicated by the solid line. The result varies by a few percent
only over our range of sea quark masses at any given $\beta$ value. In the
$\overline{MS}$ scheme at $\mu = 2\, \mbox{GeV}$ we have~\cite{Che}
$Z_S^{\overline{MS}}(2\,\mbox{GeV}) = 1.461\, Z_S^{RGI}$. At our smallest
lattice spacing, $a \approx 0.07$ fm, 
$Z_S^{\overline{MS}} \approx 0.6$, which is certainly beyond the range of
one-loop perturbation theory, tadpole-improved or not.  

Having unscrambled renormalized valence and sea quark masses, we are now able
to fit our data by the partially quenched chiral formula (\ref{chiral}) and
determine the physical quark masses from it. In the process we replace all
masses $m$ by dimensionless quantities $m r_0$. It turns out that the data are
not sensitive to logarithmic variations in the renormalized valence quark mass
(parameters $C$ and $E$). We therefore have chosen to fit our partially
quenched pseudoscalar masses by the formula
\begin{equation}
(m_{PS}\, r_0)^2 = [A + B \, m_{\rm sea}^R r_0]\, m_{\rm val}^R r_0
          + D \, (m_{\rm val}^R r_0)^2 \,.
\label{chiralr0}
\end{equation} 
In Fig.~6 we plot our data for our largest $\beta$ value. The slope of the
data depends only rather weakly on the renormalized sea quark mass. Perhaps
most of the effect is washed out by having used $r_0$ to set the scale.
The solid curve shows the result of the fit in the limit
of vanishing sea quark mass. We find good
scaling properties. The fit parameter $A$ varies by less than 5\% over our
range of $\beta$ values. 


To fix the scale $r_0$ in physical units, we extrapolate
recent dimensionless nucleon masses, $m_N r_0$, found by the CP-PACS, JLQCD
and QCDSF-UKQCD collaborations jointly to the physical pion mass,
following~\cite{Mei}. This gives the value $r_0 = 0.467$ fm, which we will use
here. A similar result was quoted in~\cite{Aubin}.
The average mass of the up and down quarks, $m_\ell = (m_u + m_d)/2$, is found
from extrapolating $m_{PS}$ to the physical $\pi^0$ mass, setting $m_{\rm
  val}^R = m_{\rm sea}^R$ in (\ref{chiralr0}). We obtain 
$m_\ell^{\overline{MS}}(2\,\mbox{GeV})\, r_0 = 0.00981(19)$, 0.00987(17),
0.00986(18) and 0.01044(19) at $\beta=5.20$, 5.25, 5.29 and 5.40,
respectively. Similarly, the strange quark mass, $m_s$, is obtained from
the lattice value of $m_{\rm val}^R$ that brings $m_{PS}$ to the physical
$K^0$ mass, while $m_{\rm sea}^R$ is kept fixed at the corresponding physical
sea quark mass $m_\ell^R$. Owing to the fact that the valence
quarks are degenerate, we then have~\cite{pqcpt} $m_s^R = 2 m_{\rm val}^R -
m_\ell^R$. This finally gives $m_s^{\overline{MS}}(2\,\mbox{GeV})\, r_0
= 0.2525(50)$, 0.2544(45), 0.2545(47) and 0.2671(48) at $\beta=5.20$, 5.25,
5.29 and 5.40, respectively.

To be able to extrapolate our results to the continuum limit, we need to know
$a/r_0$ in the chiral limit. In Fig.~7 we show our data for $r_0/a$ together
with a fit (fitting function: exponential of a polynomial in $\beta$,
$m_{\rm sea}$), and in Fig.~8 we show the light  
and strange quark masses as a function of the chirally extrapolated lattice
spacing. Because our fermionic 
action is nonperturbatively $O(a)$ improved, we expect the error due to the
finite cut-off to be at most of $O(a^2)$. A linear extrapolation in
$(a/r_0)^2$ to the continuum limit is therefore appropriate. We estimate the
systematic error on $r_0$ to be of the order of 7\%. 
We then obtain
\begin{eqnarray}
m_\ell^{\overline{MS}}(2\,\mbox{GeV}) &=& \, 4.7(2)(3)\;\mbox{MeV}
\,,\\  
m_s^{\overline{MS}}(2\,\mbox{GeV}) &=& 119(5)(8)\;\mbox{MeV}\,,
\end{eqnarray}
where the first error is statistical and the second systematic. In
particular, $m_s/m_\ell = 26(1)$, in good agreement with leading
order chiral perturbation theory. 

To summarize, we have performed a lattice calculation of the light and strange
quark masses in full QCD with $N_f=2$ flavors of light dynamical quarks. Our
calculation differs from previous calculations in several respects. We use
nonperturbatively $O(a)$ improved Wilson fermions and perform simulations at
four different couplings with $0.07 \leq a \leq 0.12$ fm, which allows an
extrapolation to the 
continuum limit. Furthermore, an entirely nonperturbative scheme of mass
renormalization, both for sea and valence quark masses, is devised, including
the effect of quark-line disconnected contributions. The identification of
renormalized sea and valence quark masses greatly facilitates the
extrapolation to the chiral limit. We believe that our method will be useful
in other applications of partially quenched chiral perturbation theory as well.

\begin{acknowledgments}
This work is supported by DFG under contract FOR 465 (Forschergruppe
Gitter-Hadronen-Ph\"anomenologie). The numerical calculations have been
performed at LRZ (M\"unchen)~\cite{Mei}, EPCC (Edinburgh)~\cite{UKQCD}, DESY 
(Zeuthen) and NIC (J\"ulich).
\end{acknowledgments}


\end{document}